\documentstyle[12pt]{article}
\begin{document}

\def\Bbb{\cal}

\def\figs{
\put(10.00,15.00){\framebox(20.00,10.00)[cc]{Source}}
\put(45.00,25.00){\framebox(10.00,5.00)[cc]{P}}
\put(45.00,10.00){\framebox(10.00,5.00)[cc]{}}
\put(30,20){\vector(1,0){17}}
\put(47,15){\line(0,1){10}}
\put(50,15){\line(0,1){10}}
\put(53,15){\line(0,1){10}}
\put(53,23){\vector(1,0){12}}
}

\hyphenation{Ger-lach po-la-ri-zer ortho-go-nal
ad-joint Gedan-ken-com-pu-ter re-pre-sen-table
tele-com-mu-ni-ca-tion satis-fi-abi-li-ty
}

\title{
Logic programming as quantum measurement
}
\author{
Roman R. Zapatrin\\
Friedmann Lab. for Theoretical Physics\\
SPb UEF, Griboyedova 30/32\\
191023, St-Petersburg, Russia
}
\maketitle
\abstract{
The emphasis is made on the juxtaposition of (quantum~theorem)
proving versus quantum (theorem~proving). The logical contents of
verification of the statements concerning quantum systems is
outlined. The Zittereingang (trembling input) principle is
introduced to enhance the resolution of predicate satisfiability
problem provided the processor is in a position to perform
operations with continuous input. A realization of Zittereingang
machine by a quantum system is suggested.
}
\medskip
\par

\centerline{\bf\large
Introduction
}
\par

What is the desirable way to broaden the facilities of processors
in order to have progress in solving NP-hard problems? The
principle of trembling input I am going to put forth in this paper
will require a flexibility of processor. That means that it will be
assumed that the processor is able to deal with "intermediate
outputs", or, in other words, that it will be possible to pass
continuously from one input to another. We shall also assume that
all possible inputs form a linear space, and that the result
depends contiuously on input preserving linear combinations.

To elucidate the idea of the principle I suggest to consider the
SAT problem. Suppose that, solving some problem, you succeded to
reduce  it  to  the following  one:  given  a propositional  form
$P(x_1,\ldots ,x_n)$  of,  say,  $n=100$ variables, you have to
check whether there exists an n-tuple $(c_1 ,\ldots ,c_n)$ such
that $P(c_1,\ldots ,c_n)='true'$.
\par
So, if there is no heuristical methods to prove the theorem
$$
\exists x_1,\ldots ,\exists x_n\quad  P(x_1 ,\ldots ,x_n)
$$

it remains the only (classical) way of solution, namely, to choose
at  random the values of the input $n$-tuples, substitute them to
the machine calculating the form $P$,  and  wait  while  the 'true'
will appear on the output.  Or, if you have $2^n$   such  machines,
do  it simultaneously. When $n$ is large enough, both possibilities
are unrealistic, and you have to confirm that there are no
classical means to  prove  this  this theorem.
\par

{\bf The  Zittereingang principle}  which I am going to  put
forward could provide the non-constructive solution of the problem.
Its basic idea is the following. Instead of preparing and
re-preparing the input register at different input states in order
to search the witness of $P=\hbox{'true'}$ we make the input state
{\it tremble}.  Then, instead of checking the output value you
measure the {\it derivative} of the output. If the value of $P$ is
always NO, the output will not tremble you will have the zero value
of the derivative. Whereas there are YESs among output values, the
derivative will be nonzero. So, checking the value of the
derivative, you can prove or reject the theorem.
\par

To realize this project, I suggest to call on
quantum effects.

% Section 1
\section{ Polarized electrons } The potential Reader of this paper
is assumed to be far away from quantum mechanics, so a brief
outline of the simplest physical experiment where quantum effects
arise seems appropriate.  This will be the famous Stern-Gerlach
experiment with polarized electrons.  The equipment needed for this
experiment is depicted at Fig.1.
\medskip

% Fig. 1
\begin{figure}
\unitlength=1.5mm
\linethickness{0.6pt}
\begin{picture}(78,30)
\put(2,0){\mbox\figs}
\put(55.5,17){\vector(1,0){12}}
\put(67,21){\framebox(4,3)[cc]{\small D}}
\put(5,20){\vector(1,0){5}}
\put(9,21){\mbox{$y$}}
\put(5,20){\vector(0,1){5}}
\put(6,24){\mbox{$x$}}
\put(5,20){\vector(-1,-1){4}}
\put(1,17){\mbox{$z$}}
\end{picture}
\caption{
The Stern-Gerlach experiment. The source (cathode) $S$
emits a beam of electrons. The polarizer $P$ splits the initial
beam into two. The detector $D$  counts the number of absorbed
electrons: whenever an electron is absorbed, the detector clicks.
}
\end{figure}
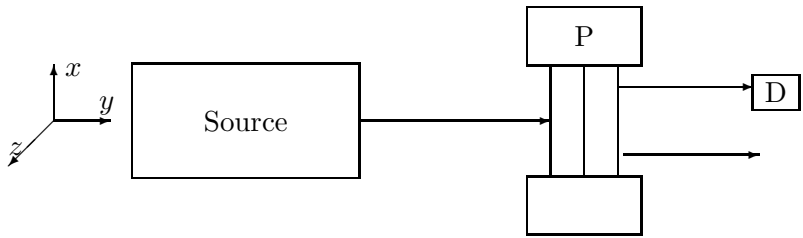

\par
There will be two possible orientations of the polarizer
(w.r.t.  the frame at the left margin of Fig.1) and two possible
positions for detector within this scheme.  So, we shall be able
to prepare four sorts of beams:
\begin{enumerate}
\item{The polarizer $P$ is $x$-oriented, and the upper
beam is chosen.}
\item{The polarizer $P$ is $z$-oriented, and the right
beam is chosen.}
\item{The polarizer $P$ is $x$-oriented, and the lower
beam is chosen.}
\item{The polarizer $P$ is $z$-oriented, and the left
beam is chosen.}
\end{enumerate}

% Fig.2
\begin{figure}
\unitlength=1mm
\linethickness{0.4pt}
\begin{picture}(30,30)
\put(15,15){\circle{14}}
\put(5,15){\vector(1,0){20}}
\put(25,13){\mbox{$z$}}
\put(8,15){\circle*{1}}
\put(6,16){\mbox{$4$}}
\put(22,15){\circle*{1}}
\put(23,16){\mbox{$2$}}
\put(15,5){\vector(0,1){20}}
\put(15,8){\circle*{1}}
\put(13,5){\mbox{$3$}}
\put(15,22){\circle*{1}}
\put(13,23){\mbox{$1$}}
\put(15,25){\mbox{$x$}}
\end{picture}
\caption{Reminder on state notation: the filled circle denotes the trace of
the appropriate beam in the $(xz)$ plane.}
\end{figure}
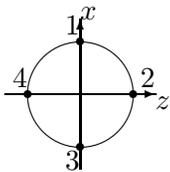

\medskip
\par
 From now one can consider the new source of electrons {\it
prepared in a definite state}(Fig. 3).

% Fig.3
\begin{figure}
\unitlength=1mm
\begin{picture}(80,40)
\figs
\put(2,2){\framebox(56,35)[cc]{}}
\end{picture}
\caption{The source of electrons prepared in a definite state.}
\end{figure}
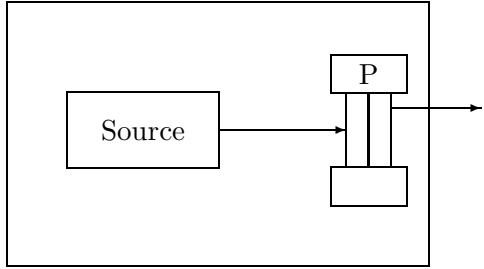

\par
The discovery of Stern and Gerlach was the following. Suppose that
the initial beam is prepared in the state 1. The making it pass
through the $x$-polarizer, one can see that the beam will not split
anymore all electrons will go up. That means that whenever an
electron is emitted in the state 1 and passes through the filter
checking "Whether the state is 1?", the answer will be YES with
certainty. Whereas, if the same beam will pass through the
$z$-polarizer, it will split into two. Thus, provided the electron
is emitted in state 1, the probability to detect it being in the
state 2 is $1/2$, and the same for the state 4 (pip into the
reminder at Fig.2 above). These results can be summarized as
follows.
\medskip
\par

Denote by $P(i,k)$ the probability for the electron to be detected
in the state $k$ provided it was emitted in the state $i$. Then
$$
P(i,k)=\cases{
1 & ,\quad$i=k$\cr
0 & ,\quad$i-k=1(\hbox{mod}2)$\cr
1/2 & otherwise
}\eqno{(1.1)}
$$
The striking property of quantum mechanics is that the results (1.1)
is ALL that we can observe by means of this equipment. It is in
principle impossible to measure a pair of properties, say, 1 and 2
simultaneously. Such series of experiments resulted in the famous
Bohr's {\bf complementarity principle}: there are physical
properties which can not be measured simultaneously. Given two such
properties $\cal A$ and $\cal B$, the measurement of $\cal A$
unavoidably affects the result of measurement of $\cal B$. This
principle is the fundamental law of the Nature: we can like it or
not, but no its violation was yet found.

% Section 2
\section{
A query model: (quantum theorem) proving
}
\medskip
\par
In this section I am going to tackle the problem of determining the
initial state of quantum objects. To make the conclusions more
transparent, a more complicated quantum system will be studied. It
will have 8 possible initial states: $1,\ldots 8$. At each moment
we shall be able to test one of these states. Like in the previous
section, we shall have the following outcomes. $P(i,k)$ here means
the probability to have the YES reply to the query $k$ provided the
initial state of the object was $i$:
$$
P(i,k)=\cases{
1 & ,\quad$i=k$\cr
0 & ,\quad$i-k=1(\hbox{mod}4)$\cr
1/4 & otherwise
}\eqno{(2.1)}
$$

This is the analog of the formula (1.1). This model can be realized
by the experiment with ions, analogous to the Stern-Gerlach one,
but the filters $x$ and $z$ will split the incoming beam into four.
The quantum theorem I am going to prove is formulated as follows.
\par
Suppose we have a source of ions, which emits them by one, and all
of them are prepared in {\it the same initial state}, but we do not
know this state. The question is formulated as:
\medskip
\par
\noindent{\bf Which is the initial state?}
\par
Denote by $q$ the query {\it Is the state $q$?} Looking at the
formula (2.1) we can see that after at most 6 steps (emitted ions)
the question reduces to the following {\it Is the state $i$ or
$k$?} where $i$ is even and $k$ is odd.  The algorithm of this
reduction is drawn at Fig.4.

% Figure 4
\begin{figure}

\unitlength=1mm
\special{em:linewidth 0.4pt}
\linethickness{0.4pt}

\begin{picture}(116,71)

\put(15,40){\circle{5}}
\put(15,42.5){\line(0,1){5}}
\put(15,50){\circle{5}}
\put(15,52.5){\line(0,1){5}}
\put(15,60){\circle{5}}
\put(17,62){\line(1,1){6}}
\put(25,70){\circle{5}}
\put(35,40){\circle{5}}
\put(35,42.5){\line(0,1){5}}
\put(35,50){\circle{5}}
\put(37,52){\line(1,1){6}}
\put(45,60){\circle{5}}
\put(43,62){\line(-2,1){15}}
\put(55,30){\circle{5}}
\put(55,32.5){\line(0,1){5}}
\put(55,40){\circle{5}}
\put(57,42){\line(1,1){6}}
\put(65,50){\circle{5}}
\put(63,52){\line(-2,1){15}}
\put(75,30){\circle{5}}
\put(77,32){\line(1,1){6}}
\put(85,40){\circle{5}}
\put(83,42){\line(-2,1){15}}
\put(95,20){\circle{5}}
\put(97,22){\line(1,1){6}}
\put(105,30){\circle{5}}
\put(103,32){\line(-2,1){15}}
\put(115,20){\circle{5}}
\put(113,22){\line(-1,1){6}}

\put(5,30){\circle*{2}}
\put(6,31){\line(1,1){7}}
\put(5,40){\circle*{2}}
\put(6,41){\line(1,1){7}}
\put(5,50){\circle*{2}}
\put(6,51){\line(1,1){7}}
\put(15,30){\circle*{2}}
\put(15,31){\line(0,1){6}}
\put(25,30){\circle*{2}}
\put(26,31){\line(1,1){7}}
\put(25,40){\circle*{2}}
\put(26,41){\line(1,1){7}}
\put(35,30){\circle*{2}}
\put(35,31){\line(0,1){6}}
\put(45,20){\circle*{2}}
\put(46,21){\line(1,1){7}}
\put(45,30){\circle*{2}}
\put(46,31){\line(1,1){7}}
\put(55,20){\circle*{2}}
\put(55,21){\line(0,1){6}}
\put(65,20){\circle*{2}}
\put(66,21){\line(1,1){7}}
\put(75,20){\circle*{2}}
\put(75,21){\line(0,1){6}}
\put(85,10){\circle*{2}}
\put(86,11){\line(1,1){7}}
\put(95,10){\circle*{2}}
\put(95,11){\line(0,1){6}}
\put(105,10){\circle*{2}}
\put(106,11){\line(1,1){7}}
\put(115,10){\circle*{2}}
\put(115,11){\line(0,1){6}}

\put(83,7){\makebox(0,0)[cc]{56}}
\put(93,7){\makebox(0,0)[cc]{58}}
\put(103,7){\makebox(0,0)[cc]{67}}
\put(113,7){\makebox(0,0)[cc]{78}}
\put(43,17){\makebox(0,0)[cc]{36}}
\put(53,17){\makebox(0,0)[cc]{38}}
\put(63,17){\makebox(0,0)[cc]{45}}
\put(73,17){\makebox(0,0)[cc]{47}}
\put(3,27){\makebox(0,0)[cc]{16}}
\put(13,27){\makebox(0,0)[cc]{18}}
\put(23,27){\makebox(0,0)[cc]{25}}
\put(33,27){\makebox(0,0)[cc]{27}}
\put(43,27){\makebox(0,0)[cc]{34}}
\put(3,37){\makebox(0,0)[cc]{14}}
\put(23,37){\makebox(0,0)[cc]{23}}
\put(3,47){\makebox(0,0)[cc]{12}}

\put(14,39){\makebox(0,0)[lb]{6}}
\put(14,49){\makebox(0,0)[lb]{4}}
\put(14,59){\makebox(0,0)[lb]{2}}
\put(24,69){\makebox(0,0)[lb]{1}}
\put(44,59){\makebox(0,0)[lb]{2}}
\put(34,49){\makebox(0,0)[lb]{3}}
\put(34,39){\makebox(0,0)[lb]{5}}
\put(64,49){\makebox(0,0)[lb]{3}}
\put(54,39){\makebox(0,0)[lb]{4}}
\put(84,39){\makebox(0,0)[lb]{4}}
\put(54,29){\makebox(0,0)[lb]{6}}
\put(74,29){\makebox(0,0)[lb]{5}}
\put(104,29){\makebox(0,0)[lb]{5}}
\put(94,19){\makebox(0,0)[lb]{6}}
\put(114,19){\makebox(0,0)[lb]{6}}

\end{picture}
\caption{The reduction algorithm. Each node with a number inside is
a test. Left edges are associated with YES outcomes, right ones are
NOT outcomes. The filled nodes correspond to the reduced questions.
For instance, 38 means {\it $3$~or~$8$}}

\end{figure}
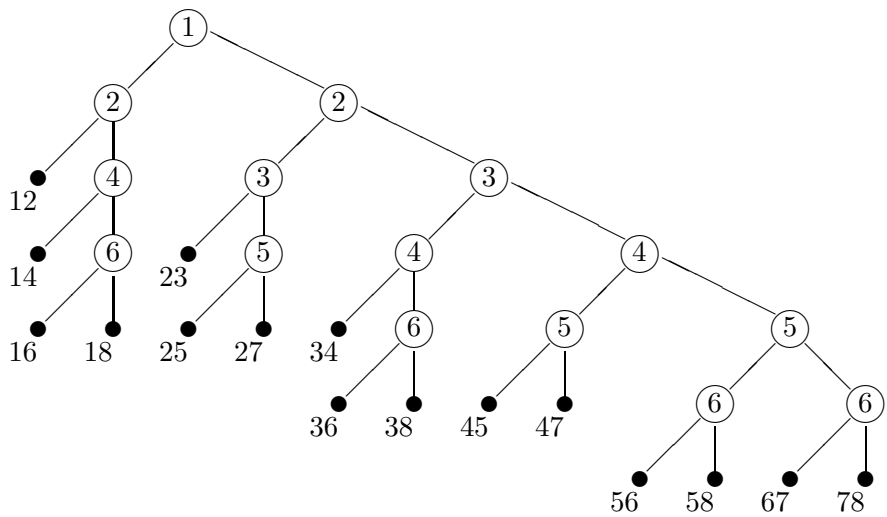

\medskip
\par
Now suppose we stopped on the alternative 12. If we then perform
the test, say, 1, the positive reply which is possible in virtue
of (2.1) will not clarify the situation in virtue of the formula
(2.1). To know the state, we must perform a test, say, 3. If the
reply is YES, nothing is clear.  Meanwhile, if NO is obtained, it
is the solution of the problem:  that means that the state is
definitely 2. Whereas we see that {\bf there is no upper bound for
the number of tests to get the definite answer} irrespective to
that the initial state was pure one!
\par

The corollary is that in quantum situation you can have a problem
of choice from a finite number of alternatives having no upper
bound for its resolution time.
\par

Note that the quasi-quantum query model can be realized by
classical automata having nothing quantum in their nature. These
are so-called {\it normalized automata} (Grib and Zapatrin, 1990).
 From the pure logical point of view normalized automata may be
thought of as a sort of "Skolemized quantum systems" grasping only
the logical structure of testable properties.

% Section 3
\section{
A piece of quantum mechanics.
}
\medskip
\par

The mathematical formalism of quantum mechanics is the steady and
rigid machinery verified by enormous number of experiments and
applications. I shall briefly introduce its piece needed in the
sequel. In quantum mechanics, a physical object is associated with
a Hilbert space ${\cal H}$, called {\it state space}. I shall
consider here only finite-dimensional case. A (pure) state of the
object is associated with a unit vector $\psi\in\cal H$. When $\cal
H$ is realized as functional space $\psi$ is just the wavefunction,
or "matter wave".
\par
The observable entities are associated with self-adjoint operators
in $\cal H$. Let $A$ be the operator associated with an observable
$\cal A$. Since $A$ is self adjoint, it can be decomposed in
accordance with the spectral theorem:
$$
A=\sum a_iP_i
$$
where $P_i$ are mutually orthogonal projectors in $\cal H$, and
$a_i\in\Bbb R$ are possible values of the observable $\cal A$. The
heart of quantum measurement theory is the
\par
\noindent
{\bf Projection postulate}: Let $\psi$ be the initial state of
the object. Then the measurement of the observable $\cal A$ in the
state $\psi$ results in:

\begin{itemize}
\item{
The value $a_i$ of $\cal A$ is obtained with the probability
equal to $\mid\mid P_i\psi\mid\mid^2$. In particular, when $\psi$
is an eigenstate of $A$, the appropriate value $a_i$ is obtained
with certainty}
\item{
The state of the object changes after the measurement to the
eigenstate associated with the observed value of $\cal A$. In
particular, when the initial state $\psi$ is the eigenstate of $A$,
the measurement does not change the state of the object.}
\end{itemize}
\medskip
\par

The new fundamental notion brought by quantum mechanics is that of
complementary measurements. In the formalism they are associated
with non-commuting operators. Due to the Bohr's principle, such
measurements are not performable simultaneously in principle.

\par
The last (but not the least) thing that will be needed is the
description of what happens with a quantum mechanical system while
one does not touch it, that is, how it {\it evolves in time}. This
is described as:
$$
\psi(t)=U(t)\psi(0)=exp(-iHt)\psi(0)
$$
where $U(t)$ is a semigroup of unitary operators
re\-pre\-sen\-table in the form $exp(-iHt)$ with a self-adjoint
operator $H$ called the {\it Hamiltonian} of the evolution. In
particular, when $\psi(0)$ is an eigenvector of $H$, the initial
state is unchanged up to a factor $\alpha(t)$ such that
$\mid\alpha(t)\mid^2=1$ and the test verifying the state $\psi(0)$
will always give the YES answer.  \par That is all from quantum
mechanics that will be needed for further purposes.

% Section 4
\section{
Classical Gedankencomputer
}
\medskip
\par
In this section section I describe the classical
computational process as quantum one.
\medskip
\par
\noindent
{\bf Loading
input data.} It is supposed that there is a  source of  physical
entities, call them {\it launches}, having the state space rich
enough to encode $2^{n}$ input values. As a matter of fact, the
state space must be broader to include the START and HALT states to
enable the reversibility of the process. The launch enters the
input register (measuring device having $n$ yes-no, or 0-1
controls) which prepares it in one of its orthogonal  pure states.
\medskip
\par
\noindent
{\bf Computation}  is  the  evolution  of  the  launch  within  the
processor controlled to that extent that before the computation
starts  the  parameters of Hamiltonian responsible for the
evolution of the launches are set  up  (in other words, the program
is loaded before the  computation).  Whereas  during the
computation  no  intermediate  measurements  are  performed.  Some
ideas towards the realization of  this  process  were  proposed  by
Deutsch (1989).
\medskip
\par
\noindent
{\bf Output} is the source of particles (they may not be the
initial launches) having only two pure states and produced by the
computer in such a  way  that their states are associated with the
calculated values.
\medskip
\par
So, the classical computational process looks as follows.
\medskip
\par
\noindent
{\bf Installation of program.} Let $F(x_{1},...,x_{n})$  be  the
function whose values are to be calculated. That
means that the evolution law for  the processor is fixed up in such
a way that each  pure  state  prepared  by  the input register
affects the emission of the output launch in one of the final
states.
\medskip
\par
\noindent
{\bf Calculation.} The values $(x_{1},...,x_{n})$ of input
variables  are  imposed  by appropriate setup of the input
controls.  Then  the  launch  is  emitted.  It passes through the
input  register  which  prepares  it  in  the  pure  state
$\psi(0)$ and enters the input  gate  of  the
processor.  By  the  end  of calculation a polarized electron is
emitted by the processor. It rushes  into the output filter which
checks whether the state is HALT (and it certainly will be so since
we have organised the evolution within the processor in such a
way).
\medskip
\par
It is essential to note that the Hamiltonian of the  processor  is
such that if the calculation with  the  same  input  data  is
repeated,  you  are compelled to get the same result. So, on
classical input data  the  described Gedankencomputer works as a
deterministic machine.

% Section 5
\section{
Uncertainty at service
}
\medskip
\par
In this section I am going to show how the Gedankencomputer
described in the previous section may be converted into Quantum
Theorem Prover.
\medskip
\par
In order to convert the computer to QTP- Quantum  Theorem
Prover,  both classical input and output registers are removed
while the processor is  kept unchanged.  {\bf The input register}
is replaced by an apparatus  measuring complementary values. The
crucial point is the following: {\it the prepared state must be the
superposition  of ALL input states with all NONZERO coefficients}.
What is needed, is to suppose that the process of  computation  of
each particular  value $P(x_{1},\ldots ,x_{n})$  is  a  quantum
process.
\medskip
\par

The launch rushes  into  the  processor,  the latter  will start
working governed by  its dynamical laws (=computes $P$ according to
the program). Finally, since the quantum evolution is linear, the
result will be the superposition of  all  classical results with
the  same coefficients.  Then, instead  of measuring the output
value, you make {\bf the output register} be an observable
complementary to the classical output.
\medskip
\par

Which should be this complementary observables? The idea is the
following: {\bf we make the input state oscillate (in time or in
space, at convenience), and measure the value of the derivative of
the output}. Then, since the processor still computes $P$, we have
two alternatives:
\begin{enumerate}
\item{If there is no $x_{1},\ldots ,x_{n})$ such that $P(x_1,\ldots
x_n)$ is true, then the value of the derivative will be zero}
\item{If there are $x_{1},\ldots ,x_{n})$ such that $P(x_1,\ldots
x_n)$ is true, then the value of the derivative will be nonzero}
\end{enumerate}
\medskip
\par

The objection against this idea is that the value of the derivative
in the case when we have, say, the only satisfying set for $P$, may
be of the order $1/2^n$. To avoid this trouble return to the
description of the classical Gedankencomputer. When it is converted
to QTP, we have some spare "degrees of freedom" to define the
"program" for the processor. Namely, we are confined by only
{\it classical outputs for classical inputs}, and, since the state
space of the launch is broader, we are free to define the evolution
for {\it non-classical} input states. This is the main issue making
QTP different from other quantum computers (for review on
quanrum computers see Bennett, 1993).
\medskip
\par

So, in principle, the satisfiability problem can be solved by QTP
at one $P$-calculation state. There is also a problem of
preparation of the superposition of all input states, however it
was shown (Deutsch and Jozsa, 1992) that it can be done for
polynomial time.
\medskip

\centerline{\bf
CONCLUDING REMARKS
}
\medskip
\par

I endeavored to show how quantum effects such as superposition of
states and wave properties of the particles can be used for
calculation  purposes.  The proposed Quantum Theorem Prover is
merely an imaginary machine.  However, suppose it may  exist,  it
can  drastically  influence  many  principles  of programming.  For
example,  such   problem   as   SAT:   satisfiability   of
propositional forms is solved on classical computers for the  time
$T\simeq \exp (n)$, where $n$ is the number of variables. Meanwhile
QTP  solves  it  for polynomial time (required for the preparation
of the input).  Accepting QTPs requires new options in programming
languages:  the assignment of value should be replaced by the
preparation of the input register.
\medskip

\centerline{\bf
ACKNOWLEDGMENTS
}
\medskip
\par
I take this opportunity to mention that the idea of QTP has
grown from my attempts to understand what are Labelled Deductive
Systems introduced  by Dov Gabbay (to appear). The valuable
comments made by Richard Jozsa are aknowledgedas well as the
telecommunication support provided by Pavlov Enterprise
(St-Petersburg).
\medskip
\centerline{\bf
REFERENCES
}
\medskip

Deutsch, D. (1989),
{\it Quantum Computational Networks},
Proceedings of  the  Royal Society,
{\bf A425}, 73

Grib A.A., Zapatrin R.R.(1990),
{\it Automata Simulating Quantum Logics},
International Journal of Theoretical Physics,
{\bf 29}, 113

Deutsch D., Jozsa R.(1992),
{\it Rapid solution of problems by quantum computation},
Proceedings of the Royal Society of London, ser. A,
{\bf 439}, 553

Gabbay, D. (to appear),
"LDS - Labelled Deductive Systems",
Oxford  University Press

Bennett, C.H.(1993),
{\it Quantum Computers: Certainty from Uncertainty},
Nature,
{\bf 362}, 694

\end{document}